\documentclass[twocolumn,prc,preprintnumbers,nofootinbib,superscriptaddress,a4]{revtex4-2}
\usepackage{epsfig,graphicx,float}
\usepackage{mathrsfs}
\usepackage{rotating}
\usepackage{url}
\usepackage{esvect}
\usepackage{subfigure}
\usepackage{multirow}
\usepackage{amsmath}
\usepackage{amssymb}
\usepackage{amsfonts}
\usepackage{bm}
\usepackage{xcolor}
\usepackage[warn]{textcomp}
\usepackage{gensymb}
\usepackage{siunitx}
\usepackage{tabularx}
\usepackage{physics}
\usepackage{comment}
\usepackage{orcidlink}
\usepackage[normalem]{ulem}

\renewcommand{\vec}[1]{\boldsymbol{#1}}
\newcolumntype{C}[1]{>{\centering\arraybackslash}X}
\usepackage{hyperref}
\begin{document}
	\title{Global Trends of Nuclear Radii and Binding Energies from the NUCLEI-PACK Framework}
	\author{H. M. Maridi\orcidlink{0000-0002-2210-9897}}
	\email[Corresponding author: 
	]{hasan.maridi@manchester.ac.uk}
	\affiliation{Department of Physics and Astronomy, The University of Manchester, Manchester M13 9PL, UK}
	\date{\today}

\begin{abstract}	
I present a proof-of-concept study of nuclear radii and binding energies within NUCLEI-PACK, a novel semi-classical framework based on optimized sphere packing of nucleons and clusters. 
In this approach, proton and neutron positions are determined through a constrained packing algorithm with uniform nucleon radii, followed by global optimization across the nuclear chart ($1 \leq A \leq 250$). From these geometric configurations, I compute charge and matter radii, as well as binding energies. The model successfully reproduces global trends in nuclear size and stability, while maintaining transparency and computational efficiency. A fit of the Coulomb and surface terms of the semi-empirical mass formula shows remarkable agreement with theoretical expectations. These results establish the feasibility of NUCLEI-PACK as a systematic tool for describing nuclear bulk properties across the chart of nuclides.
\end{abstract}
\maketitle	
\section{Introduction}	
The nuclear many-body problem remains one of the most complex challenges in modern physics. While macroscopic models, such as the Liquid Drop Model~\cite{Bethe1936,Meitner1939}, provide valuable intuition, and large-scale quantum calculations achieve high precision~\cite{Carlson2015}, there is a pressing need for a computationally tractable, microscopic framework that offers deep physical insight. This work introduces a novel semi-classical model that constructs nuclei from the ground up by assigning spatial coordinates to each constituent proton and neutron. The foundation of this framework lies in nucleon configurations derived from optimized sphere-packing algorithms. This approach is physically motivated by the strong short-range repulsion of the nuclear force, which prevents nucleons from overlapping and effectively models them as hard spheres at close distances~\cite{Reid1968}.

Conceptually, this model aligns with a rich history of geometric approaches that view nuclei as structured assemblies rather than uniform drops. Meitner and Frisch~\cite{Meitner1939}, in their famous 1939 explanation of nuclear fission, were among the first to emphasize the classical analogy between nucleons in atomic nuclei and the random close packing of hard spheres. Pauling’s close-packed spheron theory, for instance, envisioned nuclei as dense clusters of stable subunits~\cite{Pauling1965a,Pauling1965b}, while the face-centered-cubic (FCC) lattice theory proposed ordered crystalline arrangements of nucleons~\cite{Coo87}. More recently, advances in the study of random close packing (RCP) have further strengthened this perspective, with packing arguments being applied to estimate nucleon densities in neutron stars~\cite{Bra23}, infer effective nucleon sizes~\cite{Kai24}, and explain neutron enrichment in stable nuclei via binary-mixture RCP theory~\cite{Zac22,Anzivino2024}.

Building on these motivations, the present framework develops packing-based ideas into a flexible computational laboratory for nuclear structure. Preliminary applications demonstrate that this framework not only captures essential features of binding and geometry, but also offers predictive power for a wide range of nuclei. The goal of this paper is to provide a proof-of-concept by applying this framework to model the global trends of charge and matter radii, as well as the binding energies, of stable and long-lived nuclei. This geometric perspective, coupled with computational tractability, provides a promising avenue for connecting microscopic nucleon arrangements with emergent nuclear properties.

\section{Theoretical Framework}
\subsection{Background on the Sphere Packing Problem}
The Sphere Packing Problem (SPP) is a classical optimization problem that seeks an arrangement of non-overlapping spheres within a bounded or unbounded region such that the packing density is maximized. Beyond its mathematical elegance, SPP has important applications in physics, chemistry, and materials science, see for example Ref. \cite{Zhou2024} and therein.
Algorithms for solving the Sphere Packing Problem (SPP) generally fall into two categories: 
(i) random sphere packing, often used for simulations in materials science and particle systems, which easily scale to thousands of particles but achieve modest packing densities; and 
(ii) dense sphere packing, which treats SPP as an optimization problem and aims to achieve the densest possible arrangements.
 A particularly relevant variant for my work is the Packing Equal Spheres in a Sphere (PESS) problem~\cite{Specht,Huang2011,Huang2012,Hallah2013,Hifi2017,Hifi2018}. Given $n$ identical spheres, the objective is to fit them into the smallest possible enclosing sphere. For small $n \leq 23$, most optimal packings are obtained analytically, while for larger $n$, benchmark solutions are documented by Specht using unpublished algorithms. This variant is directly relevant to this work, where the nucleons are treated as hard spheres packed into the spherical volume of the nucleus. The present work utilizes a computational framework based on the PESS problem to generate the spatial coordinates of each proton and neutron.

\subsubsection{Problem Formulation and Elastic Model}
The PESS problem can be formulated as a non-linear constrained optimization problem, where the goal is to minimize the radius of the spherical container ($R$). First, all nucleon spheres must be contained within the larger container sphere:
\begin{equation}\label{SPM1}
	\|\vec{p}_i\| \leq (R - r_0), \quad \quad 1 \leq i \leq n
\end{equation}
Second, the nucleon spheres must not overlap with each other:
\begin{equation}\label{SPM2}
	\|\vec{p}_i - \vec{p}_j\| \geq 2 r_0, \quad \quad 1 \leq i, j \leq n, \ i \neq j
\end{equation}

where $r_0$ is the radius of the small spheres and $\vec{p}_i$ is the position vector of the center of the $i$-th sphere.

These non-linear constraints are typically solved using a relaxed model, such as the elastic model \cite{Huang2011,Zhou2024}, which converts the problem into an unconstrained optimization by minimizing a total "elastic energy" that quantifies the degree of sphere overlap. The solutions employed in this work are the best-known numerical solutions derived from this optimization method.

\subsection{Nucleon Assignment}
\begin{figure}[h]
	\centering
	\includegraphics[width=0.23\textwidth]{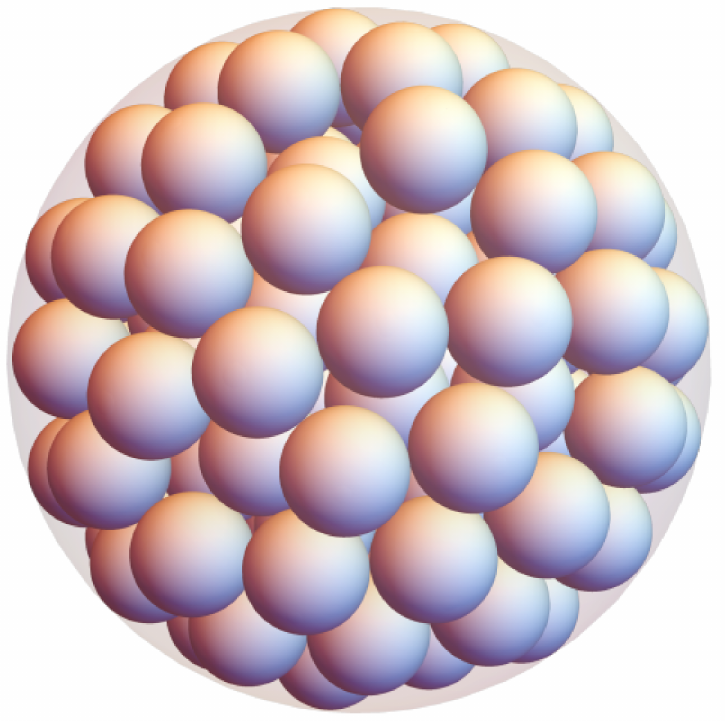}
	\includegraphics[width=0.23\textwidth]{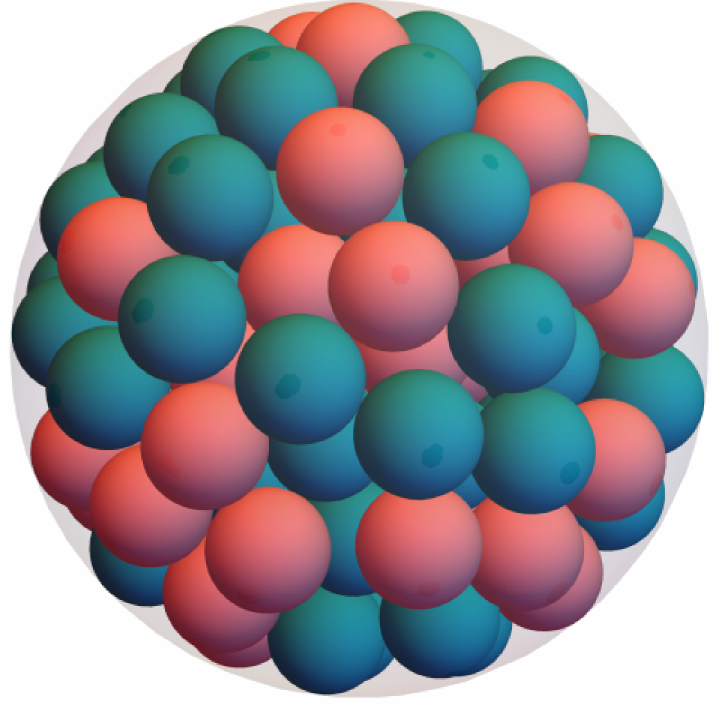}
	\caption{Sphere packing configuration for $^{120}$Sn. (a) The optimal geometric arrangement of $A=120$ nucleons from the database \cite{Specht}. (b) The result of my nucleon assignment algorithm, with protons (red) and neutrons (green) distributed at evenly spaced radial intervals.}
		\label{fig:Arrange}
	\end{figure}

I initialise nuclear geometries from best-known solutions of the \emph{Packing Equal Spheres in a Sphere} (PESS) benchmark: for each $n=A$ I take the densest known configuration of $A$ equal spheres inside a minimal enclosing sphere (Packomania dataset)~\cite{Specht}. 
Coordinates are rescaled so that all nucleons have radius $r_0$ (here $r_p=r_n=r_0$), yielding a set of centres $\{\mathbf{r}_i\}_{i=1}^{A}$ and a container radius $R_A$.
Given a nucleus with mass number $A$ and atomic number $Z$, I use the pre-calculated coordinates for an $A$-nucleon system from a public PESS database \cite{Specht}. To assign the positions of protons and neutrons, I first calculate the distance of each of the $A$ nucleons from the center of the nucleus. The nucleons are then ranked according to this distance. This approach is physically motivated by the need to more uniformly distribute the Coulomb repulsion energy throughout the nuclear volume. I assign the $Z$ protons to evenly spaced intervals within this radially sorted list to avoid spurious surface or central clustering. The remaining $N=A-Z$ nucleons are assigned as neutrons. The specific position of the $i$-th proton in the radially sorted list is given by:
\begin{equation}
	\label{eq:proton_assignment}
	\text{position}_i \;=\; 
	\left\lfloor \frac{\bigl(i - \tfrac{1}{2}\bigr)A}{Z} \;+\; \tfrac{1}{2} \right\rfloor,
	\quad i = 1,2,\dots,Z .
\end{equation}
where $i \in \{1, 2, ..., Z\}$. The process of assigning protons and neutrons to the packing configuration is illustrated in Figure~\ref{fig:Arrange} for $^{120}$Sn nucleus.

To build actual nuclei, I select long-lived/stable $(A,Z)$ pairs by thresholding experimental half-lives larger than $10^9$ s. 


\subsubsection{Future Directions: Binary Sphere Packing}
The current work utilizes a simplified equal-sphere packing model, which assumes that protons and neutrons have the same radius. A significant and ambitious future direction is to develop a new dataset based on binary sphere packing algorithms. This would allow us to use distinct radii for protons and neutrons, providing a more realistic and powerful geometric basis for the model. This is a computationally intensive task, especially for heavy nuclei, which would require the use of supercomputing facilities. Building this new dataset is a key step towards refining my framework and will enable a more fundamental study of nuclear structure, as will be discussed in my future work.
\subsection{Calculation of Nuclear Radii}
\subsubsection{Calculation of the Charge Radius}
Within my coordinate-based model, the nuclear radii are calculated from the nucleon positions. The input coordinates for the nucleons are obtained from the PESS database \cite{Specht}, which provides positions for spheres packed inside a sphere with a radius normalized to unity ($R=1$). To use these coordinates in my physical model, I scale them by an appropriate factor so that the nucleon spheres have a radius of $r_0=1.0$ fm. Coordinates are then recentered so that the nuclear center of mass lies at the origin. The RMS charge radius ($R_c$) is defined as the root-mean-square distance of the proton charge distribution from the nuclear center. In my framework, I calculate $R_c$ using two different sub-models: a point-proton model and a Gaussian-smeared model. 

The RMS charge radius in the point model is given by 
\begin{equation} \label{eq:Rc-point}
	R_c^{\text{(point)}} = 
	\sqrt{ \frac{1}{Z}\sum_{i=1}^{Z} \lVert \mathbf{r}_i^{(p)} \rVert^{2}
		+ \left\langle R_{p}^2 \right\rangle
		+ \frac{N}{Z}\, \left\langle R_{n}^2 \right\rangle
		+ \Delta_{\text{DF}} },
\end{equation}
where 
\(\mathbf{r}_i^{(p)}\) denote the proton position vectors relative to the nuclear center,  
$\left\langle R_{p}^2 \right\rangle$ and $\left\langle R_{n}^2 \right\rangle$  are the intrinsic charge radii of the proton and neutron, respectively, and 
\(\Delta_{\text{DF}}=\frac{3}{4 M^2} \) is the Darwin--Foldy relativistic correction term. 

As with the matter radius, the finite spatial spread of nucleon densities can be incorporated by smearing each proton with an isotropic Gaussian of width $\sigma_p=0.85$. This leads to
\begin{equation}\label{eq:Rc-smeared}
	R_c^{(\sigma)} \;=\; \sqrt{ \left(R_c^{\text{(point)}}\right)^{2} \;+\; 3\sigma_p^{2} },
\end{equation}
reflecting the additional contribution of the Gaussian variance in three spatial dimensions.
I use the following physical constants from Ref. \cite{Tanihata2013}: the intrinsic proton charge radius squared,$\left\langle R_{p}^2 \right\rangle = (0.877 \text{ fm})^2$, the intrinsic neutron charge radius squared, $\left\langle R_{n}^2 \right\rangle = -0.1161 \text{ fm}^2$, and the Darwin--Foldy relativistic correction, $\Delta_{\text{DF}} = 0.033 \text{ fm}^2$.

\subsubsection{Calculation of the RMS Matter Radius}
The root-mean-square (RMS) matter radius provides a fundamental measure of the spatial extent of the nucleus. 
Within the coordinate-based framework, two complementary definitions are employed: a point-nucleon version and a smeared version incorporating nucleon density distributions. 

The RMS matter radius in the point-nucleon approximation is given by
\begin{equation} \label{eq:Rm-point}
	R_m^{\text{(point)}} = \sqrt{ 
		\frac{1}{A} \sum_{i=1}^{A} \lVert \mathbf{r}_i \rVert^2
		+ \frac{Z \, \left\langle R_{m,p}^2 \right\rangle + N \, \left\langle R_{m,n}^2 \right\rangle}{A} 
	},
\end{equation}
where 
\( A = Z + N \) is the mass number, 
\(\mathbf{r}_i\) are the nucleon position vectors relative to the nuclear center, 
and $\left\langle R_{m,p}^2 \right\rangle$ and $\left\langle R_{m,n}^2 \right\rangle$ denote the intrinsic matter radii of the proton and neutron, respectively. 

To account for the finite spatial extent of nucleon wave packets, each nucleon position can be smeared with an isotropic Gaussian density of width $\sigma_p=\sigma_n=0.85$. 
In this case, the RMS radius becomes
\begin{equation}\label{eq:Rm-smeared}
	R_m^{(\sigma)}  \;=\; \sqrt{ \left(R_m^{\text{(point)}}\right)^{2}
		+ \frac{3}{A}\!\left(Z\,\sigma_p^{2}+N\,\sigma_n^{2}\right) }.
\end{equation}	
\subsection{Calculation of Binding Energy}
To calculate the total binding energy of the nucleus, I employ a mean-field approach in which
each nucleon moves in an average potential generated by all others. The central part of the interaction
is modeled using volume and surface Woods--Saxon terms:
\begin{equation}
	V_{\text{vol}}(r) = \frac{-V_r}{1 + \exp\left(\frac{r - R_v}{a_v}\right)},
\end{equation}

\begin{equation}
	V_{\text{surf}}(r) = 4a_s V_s
	\frac{\exp\left(\frac{r - R_s}{a_s}\right)}
	{\left[1 + \exp\left(\frac{r - R_s}{a_s}\right)\right]^2},
\end{equation}
with $R_i = r_i A^{1/3}$ ($i=v,s$). The total mean-field energy is

\begin{equation}
	E_{\text{mean-field}} = \sum_{j=1}^A \left[ V_{\text{vol}}(r_j) + V_{\text{surf}}(r_j) \right],
\end{equation}
where $r_j$ is the distance of the $j$th nucleon from the nuclear center. The Coulomb interaction is
still treated as a two-body sum over all proton pairs:

\begin{equation}
	E_{\text{Coulomb}} = \sum_{i < j}^{Z} \frac{e^2}{|\vec{r}_i - \vec{r}_j|}.
\end{equation}
Then, the binding energy per nucleon is
\begin{equation}\label{eq:BE-mean-field}
	\frac{B}{A} = -\frac{E_{\text{total}}}{A}
	= -\frac{E_{\text{mean-field}} + E_{\text{Coulomb}}}{A}.
\end{equation}

The model parameters are fitted to experimental binding-energy data.  

A natural future extension of this framework is to replace the mean-field potential with
direct two-body nucleon–nucleon interactions, such as Reid soft-core \cite{Reid1968} or modern chiral NN potentials.
This would provide a more microscopic description of binding energies, but requires
computational resources beyond the scope of this proof-of-concept.

\subsection{Validation of Macroscopic Terms}
To validate the physical realism of my semi-classical sphere-packing framework, I demonstrate its ability to reproduce two key components of the Semi-Empirical Mass Formula (SEMF), a well-established macroscopic model of nuclear binding energy. The SEMF describes the total binding energy ($B$) as a sum of five terms:
\begin{align}\label{eq:SEMF}
	B(A,Z) = a_{\text{vol}}A - a_{\text{surf}}A^{2/3} - a_{\text{Coul}} \frac{Z(Z-1)}{A^{1/3}} \nonumber \\ - a_{\text{sym}} \frac{(N-Z)^2}{A} + a_p\delta A^{-1/2}
\end{align}
where the second and third terms represent the surface and Coulomb contributions, respectively. This model provides a direct microscopic interpretation of these two terms.
\subsubsection{Fit of the Coulomb Term}
A crucial test of any microscopic model is its ability to reproduce well-established macroscopic behavior. To validate the geometric distribution of protons within my sphere-packing framework, I performed a fit to isolate the Coulomb energy component. The total Coulomb potential energy in my model is calculated by summing the two-body potential over all unique proton pairs:
\begin{equation}
	E_{\text{Coulomb, Micro}} = C \sum_{i<j}^{\text{protons}} \frac{e^2}{r_{ij}} ,
\end{equation}
where $r_{ij}$ is the distance between each proton pair in fm, and $C$ is a dimensionless free parameter scaling the electrostatic interaction.

This microscopic sum was then compared with the macroscopic Coulomb energy from the semi-empirical mass formula (SEMF), which models the nucleus as a uniformly charged sphere:
\begin{equation}
	E_{\text{Coulomb, Macro}} = a_c \frac{Z(Z-1)}{A^{1/3}} ,
\end{equation}
where $a_c \approx 0.714~\text{MeV}$.

\begin{figure}[tbh]
	\centering
	\includegraphics[width=0.48\textwidth]{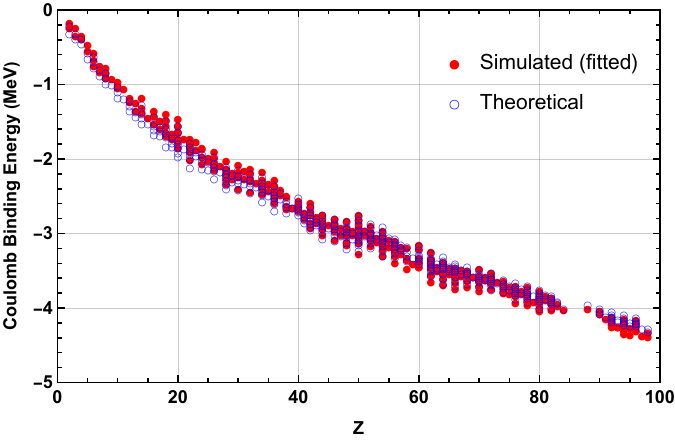}
	\caption{Fit of the microscopic Coulomb potential from this work to the macroscopic Coulomb term of the Semi-Empirical Mass Formula (SEMF). The best-fit scaling factor, $C \approx 1$, is remarkably close to unity, validating the model's ability to reproduce macroscopic behavior.}
	\label{fig:Coulomb_fit}
\end{figure}

The best-fit scaling factor was found to be $C = 0.9955$, which is remarkably close to unity. As shown in Figure~\ref{fig:Coulomb_fit}, this result demonstrates that the explicit proton distributions obtained from the packing algorithm reproduce the expected macroscopic Coulomb behavior without artificial tuning, providing strong validation of the physical realism of the model.

\subsubsection{Surface Term of the Binding Energy}
The surface contribution to the binding energy arises because nucleons on the nuclear surface
experience fewer attractive neighbors than those in the interior. Within the packing model, we
quantify this by counting \emph{under-coordinated} nucleons. A nucleon is classified as a surface
nucleon if its number of neighbors within a fixed neighbor radius $r_{\mathrm{nbr}}=3.0~\mathrm{fm}$
is smaller than a threshold coordination number $N_c=11$. The surface energy is then taken to be
proportional to the number of these nucleons:
\begin{equation}\label{eq:surface}
	E_{\text{surface}} = a_n N_{\text{surface}},
\end{equation}
where $a_n$ is the effective energy penalty per surface nucleon.

\begin{figure}[tbh]
	\centering
	\includegraphics[width=0.48\textwidth]{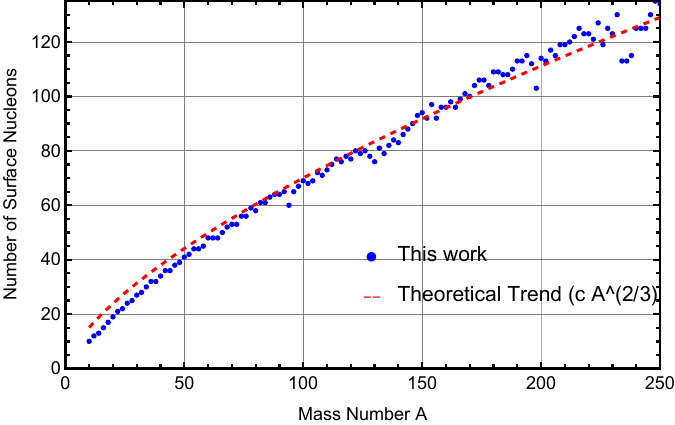}
	\caption{Number of surface nucleons identified from the packing configurations compared with the expected $A^{2/3}$ scaling law. The agreement provides a direct microscopic interpretation of the surface term in the semi-empirical mass formula.}
	\label{fig:surface_nucleons}
\end{figure}

The number of surface nucleons extracted from the packing model follows the expected $A^{2/3}$
scaling (Fig.~\ref{fig:surface_nucleons}), and can be fitted as
\begin{equation}
	N_{\text{surface}} \approx c\,A^{2/3}, \quad c = 3.25.
\end{equation}
Substituting this into Eq.~\eqref{eq:surface} gives
\begin{equation}
	E_{\text{surface}} \approx a_n c\,A^{2/3}.
\end{equation}
Comparison with the semi-empirical mass formula,
$E_{\text{surface}}^{\text{(SEMF)}} = a_{\text{surf}} A^{2/3}$,
implies the relation
\begin{equation}
	a_{\text{surf}} \approx a_n c .
\end{equation}
Using the standard SEMF value $a_{\text{surf}}\approx 17~\text{MeV}$,
we obtain
\begin{equation}
	a_n \approx \frac{17}{3.25} \simeq 5.2~\text{MeV}.
\end{equation}
Thus, each surface nucleon carries an effective penalty of about $5.2~\text{MeV}$ within this microscopic framework, while the overall $A^{2/3}$ scaling is naturally enforced by the packing geometry.

\section{Results}
\subsection{Nuclear Radii}
A first validation of the sphere-packing framework is its ability to reproduce the global systematics of nuclear sizes. Both charge and matter radii are calculated directly from the explicit nucleon coordinates, with Gaussian smearing applied to represent the finite density distribution of each nucleon.
\subsubsection{Charge Radii}
\begin{figure}[h!]
	\centering
	\includegraphics[width=0.48\textwidth]{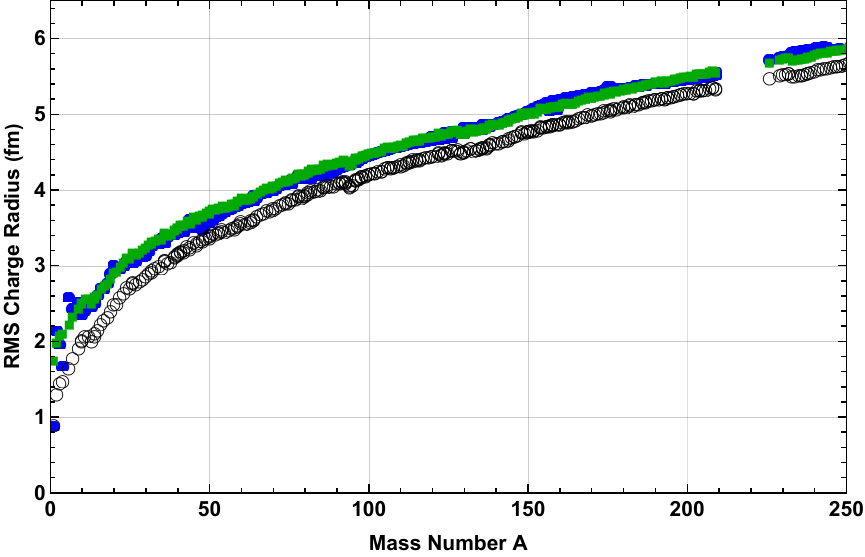}
	\caption{Calculated rms charge radii from the point model (open black circles) and the smeared model (green circles) are compared with experimental data (blue circles) from \cite{Angeli2013}. The smeared model accurately reproduces the global systematics of charge radii across the nuclear chart.}
	\label{fig:Charge_radii}
\end{figure}

A primary validation of the NUCLEI-PACK framework is its ability to reproduce the global systematics of nuclear sizes. In Figure~\ref{fig:Charge_radii}, the calculated rms charge radii (using Eqs \ref{eq:Rc-point} and \ref{eq:Rc-smeared}) are compared with experimental data across the nuclear chart. The agreement is excellent: the model captures both the overall $A^{1/3}$ scaling and the fine systematic variations between isotopes. The inclusion of the smearing correction is crucial to reproduce the experimental data \cite{Angeli2013}.

\subsubsection{Matter Radii}
\begin{figure}[h!]
	\centering
	\includegraphics[width=0.48\textwidth]{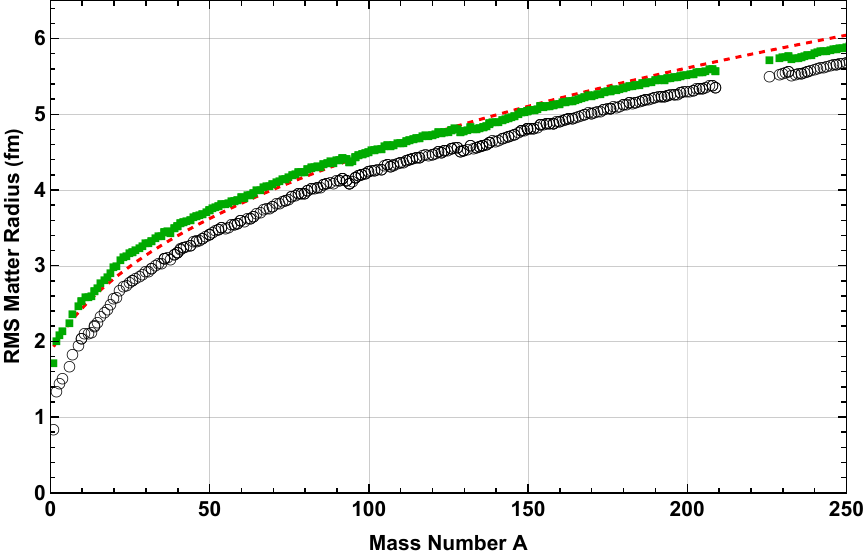}
	\caption{Calculated rms matter radii from the point model (open black circles) and the smeared model (green circles) are compared with the Fermi formula (red dashed line).}
	\label{fig:rm}
\end{figure}

The rms matter radii obtained from this framework (using Eqs \ref{eq:Rm-point} and \ref{eq:Rm-smeared}) are shown in Figure~\ref{fig:rm} and compared with the empirical Fermi formula (red dashed line). The Fermi formula for the RMS matter radius is given by:
\begin{equation}\label{eq:Rm-Fermi}
	R_{\text{Fermi}} = \sqrt{\frac{3}{5} \left(c^2 + \frac{7}{3} \pi^2 a^2\right)},
\end{equation}
where the constants are taken as $a=0.51$ fm, and the half-density radius is $c = 1.31 A^{1/3} - 0.84$ fm.

The calculated results are shown to closely follow the expected scaling, which validates the physical realism of the packing geometry. Not only is the correct global dependence provided by the model, but a microscopic interpretation of the nuclear radius in terms of nucleon configurations is also offered.

\subsection{Binding Energy}
The total binding energy per nucleon can be obtained by summing the contributions from volume, surface, and Coulomb energies within the packing framework. The explicit nucleon coordinates $\{\vec{r}_i\}$ provide the microscopic basis, while the strong interaction is modeled here by an $A$-dependent mean-field potential (Woods--Saxon). The Coulomb interaction is treated separately as a two-body force. 

\begin{figure}[tbh]
	\centering
	\includegraphics[width=0.48\textwidth]{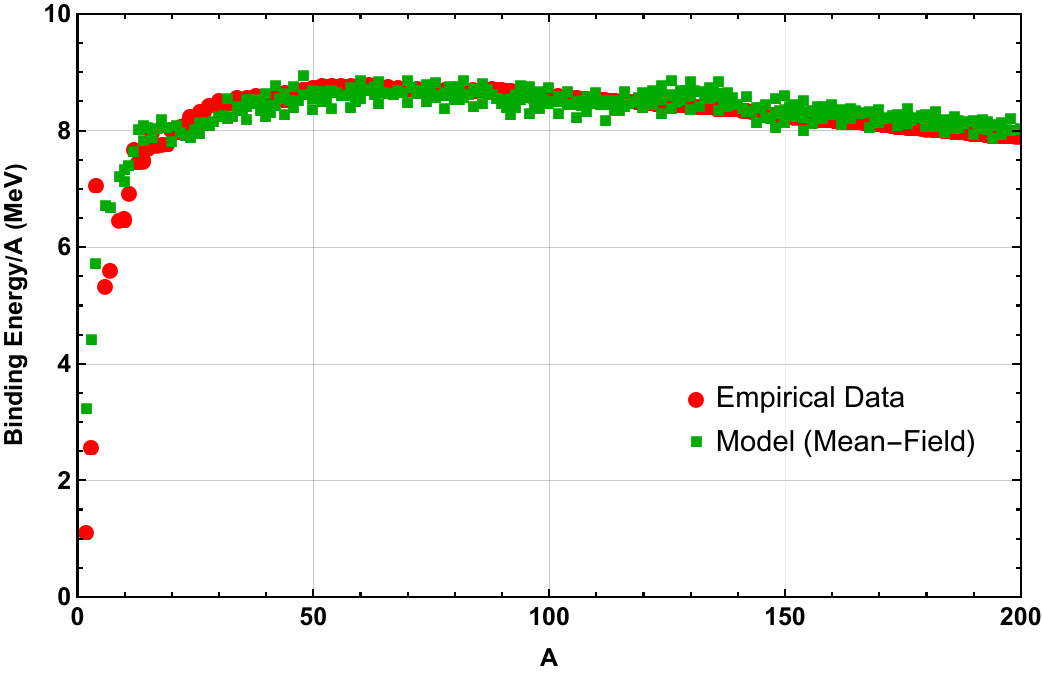}
	\caption{Calculated binding energy per nucleon from the Woods--Saxon mean-field potential (green squares) compared with experimental data (blue circles) from the NNDC \cite{NNDC}. The saturation trend is reproduced, with a maximum near mid-mass nuclei.}
	\label{fig:MeanField}
\end{figure}

Figure~\ref{fig:MeanField} shows the calculated $BE/A$ using the Woods--Saxon mean-field potential using Eq. (\ref{eq:BE-mean-field}). The parameters were determined through a quick fit to the global data, yielding the values: $V_r=13.35$ MeV, $r_v=1.2$ fm, $a_v=0.6$ fm, $V_s=4.5$ MeV, $r_s=2.0$ fm, and $a_s=1.2$ fm. The results are compared with experimental data from the NNDC \cite{NNDC}, and they reproduce the expected trend of nuclear saturation, peaking near iron and decreasing for both lighter and heavier systems. This demonstrates that even within the semi-classical framework, the essential systematics of nuclear binding emerge naturally from the geometry.

\section{Conclusion}
This work introduces a novel semi-classical framework for nuclear structure, rooted in the optimized sphere packing of nucleons. By assigning explicit spatial coordinates to protons and neutrons, the model establishes a direct connection between nuclear geometry and measurable observables, offering a new computationally tractable laboratory for nuclear structure studies.

As a proof-of-concept, I showed that the model reproduces the global systematics of charge and matter radii and naturally recovers the Coulomb and surface contributions of the semi-empirical mass formula with remarkable precision. Binding energies can be addressed through both mean-field and direct interaction approaches, while preliminary applications demonstrate the emergence of extended halo geometries and $\alpha$-cluster configurations.

Future developments will focus on incorporating essential quantum effects, such as pairing and spin–orbit coupling, and on leveraging machine learning techniques for large-scale predictions across the nuclear chart. By bridging macroscopic intuition with microscopic detail, the sphere-packing framework provides a powerful and versatile methodology, with potential applications extending from neutron-skin thickness and the $r$-process to the composition of neutron stars.



\end{document}